\documentclass[useAMS,usenatbib,usenatNishiNishibib,usegraphicx]{mn2e}
\usepackage{graphicx}

\newcommand{\pr}{$^{\prime}$}
\newcommand{\hi}{H~{\sc i} }
\newcommand{\hii}{H~{\sc ii} }

\newcommand{\rsun}{$R_\odot$}

\newcommand{\dg}{$^{\circ}$}

\title[Radial Metallicity Distribution Breaks and the Corotation]{Radial Metallicity Distribution Breaks at Corotation Radius in Spiral Galaxies}

\author[Scarano Jr. \& L\'{e}pine.]{S. Scarano Jr $^{1,2}$ \thanks{E-mail: scarano@astro.iag.usp.br /
scaranojr@ctio.noao.edu} \& J. R. D. L\'{e}pine $^1$\\
$^1$ Instituto de Astronomia, Geof\'{\i}sica e Ci\^encias Atmosf\'ericas da
Universidade de S\~ao Paulo,\\ Cidade Universit\'aria, CEP: 05508-900, S\~ao Paulo, SP, Brazil \\
$^2$ Southern Astrophysical Research Telescope (SOAR), Casilla 603,
La Serena, Chile}

\begin{document}
\maketitle

\begin{abstract}

We analyzed the relation between the corotation radii and the galactic radii at which breaks or changes of slope of the metallicity gradients occur in spiral galaxies. With this purpose we compiled the results from the literature on rotation curves, corotation radii and radial metallicity distributions of 27 galaxies, of which 16 were considered qualified to be studied in the context of this work. We re-scaled all references of each galaxy to a same framework in order to compare the results and to identify the radii where breaks and changes of slopes are found, when non-linear models fit the radial metallicities better than a linear model. In most galaxies we have found minima and breaks in radial metallicity near the corotation radius, revealing a significant correlation between these two radii, as it occurs in our Galaxy. The results are interpreted as a consequence of long-lived spiral structures, in which the star-formation rate depends on the distance to the corotation radius, producing secular effects in the observed radial metallicity distributions.

\end{abstract}


\section{Introduction}

\par The mechanisms that regulate the star-formation rate and the building up of the chemical abundances in disks of galaxies, and in particular, the role of the spiral structure, are not yet fully understood. In the past, most studies of radial metallicity distributions, especially of the oxygen abundance distribution, revealed monotonically decreasing radial gradients \citep*[eg.][]{VilaCosta92, Kennicutt93, Zaritsky94}. This simple behaviour of metallicity gradients was easily reproduced by chemical evolution models in which the most important parameters are supposed to vary smoothly with radius \citep*[see eg.][]{Molla96, Chiappini1997}. For these reasons the linear behaviour of the gradients (in logarithmic scale) seemed to be universal. However, the presence of breaks in the oxygen abundance slopes had already been observed in a number of galaxies in the decade of 1990, for instance in M101 (NGC5457) \citep{Zaritsky92, VilaCosta92}, NGC5194 \citep{VilaCosta92}, NGC 3319 and M83 (NGC 5236) \citep{Zaritsky92}, NGC3359 \citep*{Martin95}, NGC 1365 \citep{RoyWalsh97}. The real existence of these breaks was contended by \cite*{Pilyugin03}, who claimed they are due to a systematic error involving the excitation parameter, which would overestimate the oxygen abundance  in low excitation \hii regions. But at least in a few cases, like M83 \citep{Bresolin09}, NGC3359 by \cite*{Zahid2011}, NGC 4625 \citep{Goddard2011}, new studies or  re-examination of the gradients with new metallicity diagnostics showed that the breaks indeed exist.

In general the breaks in abundance gradients are attributed to variations of gas density, the presence of a bar, or coincidence with the corotation radius ($R_{CR}$).  In a recent work, \citet*{Scarano11} investigated 3 galaxies for which the $R_{CR}$ was supposed to be in the visible part of the spiral structure. They found evidences for variations of gradient slopes, and argued that they are associated with the corotation resonance.

In the case of our Galaxy a long debate concerning the existence of breaks in the gradient also occurred. After many works reporting linear gradients \citet*{Twarog1997}, based on a sample of Open Clusters, were the first to show that there is an abrupt step down of 0.3 dex in the [Fe] abundance, at about 1 kpc beyond the orbit of the Sun. However, several authors, analyzing similar samples of clusters, did not confirm the existence of the step \citep[eg.][]{Rolleston2000}. In parallel \citet{Andrievsky2002a, Andrievsky2002b, Andrievsky2002c}, performed abundance measurements of several elements in Cepheids, and found that the metallicity gradient presents a steep slope around 5-6 kpc from the center, a plateau from about 6.5 kpc to 10 kpc, followed by  a new  decreasing gradient at about 10 kpc (quoting here the radii given in the original papers). \citet{L'epine2011}, reanalyzing the most recent data on Open Clusters and Cepheids, confirmed the existence of a sharp step in the abundance gradient, and showed that it coincides with $R_{CR}$, which is now accurately known in our Galaxy. The theoretical explanation proposed for the metallicity step in our galaxy is the "pumping out" effect of corotation, which produces gas flows in opposite directions on the two sides of the resonance. This originates the ring-shaped gap in the distribution of gas observed by \citet*{Amores2009} that isolates the inner and outer sides of the gap one from the other. As a consequence, the metallicity has evolved in an independent way on the two sides. A recent study of NGC1365 \citep*{Elmegreen2009} observed the flow of gas moving out from corotation, in opposite directions on each side.

The existence of plateaux, sharp steps, or rising metallicity gradients in the outer regions of galaxies, should not be regarded as minor effects that are superimposed on a general, well understood, linear behaviour of metallicity gradients. These features reveal that important processes like radial gas flows do exist and should be taken into account in chemical evolution models.

In order to check if the same effect observed at the corotation resonance in our Galaxy plays a major role in determining the shape of the metallicity gradients in other spiral galaxies, we analyzed the data from more than a hundred references on metallicity distributions, rotation curves and corotation radii (or their equivalent spiral pattern speeds) to verify the existence of a correlation between the radius of corotation and the radius where changes in slopes, or even of discontinuities, in the metallicity gradients occurs. In Section 2 we examine the case of our Galaxy, with the purpose to establish the strategy to locate minima, points of inflexions and/or discontinuities. In Section 3 we introduce the data compiled from the literature and the procedures to standardize the results from different references. In Section 4, we explain the method used to locate the breaks, minima or inflexions in the metallicity distributions. In section 5 we present the correlation between the metallicity breaks radii and corotation radii. The remarks on individual galaxies are collected in Section 6. We conduct the discussions in Section 7, and finally, in Section 8 we summarize the main results and discuss some of the consequences of the present work.

\begin{figure}
\includegraphics[width=85mm]{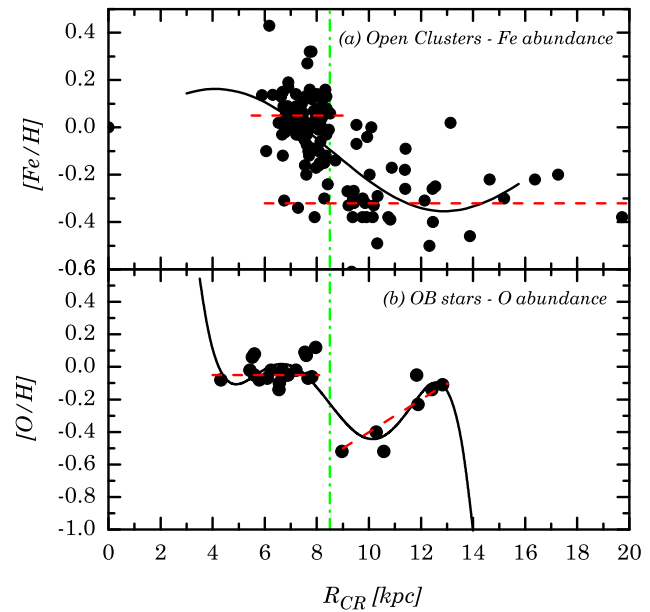} \caption {Two examples of abundance gradients in the Galactic disk, from the literature: (a) The [Fe/H] distribution of Open Clusters (Lepine et al. 2011), (b) The [O] abundance of OB stars from Daflon et al.(2004). In Panel (b) the galactocentric radii of the sources where re-calculated for \rsun = 7.5 kpc. The oxygen abundance scale was normalized to the solar abundance using [O/H]$_\odot$ = -8.6. In the two panels the dashed lines are fits of two lines (see text), the vertical line indicates the position of corotation, and the thin solid curve the fit of the data by a polynomial with the degree that best fits the observations.}
\label{fig1}
\end{figure}

\section{The abundance gradient discontinuity in our Galaxy}

\par It is important to fully understand the connection between metallicity features and corotation in our Galaxy before proceeding to the analysis of other galaxies. The corotation radius of our Galaxy is accurately known. Many papers in the past pointed that corotation was situated close to the Sun (\citealt*{Marochnik1972}, \citealt*{Creze1973},\citealt*{Mishurov1999}, among others). More recently a few direct methods not only confirmed these previous estimates but also allowed to decrease the uncertainty on the results to a few hundred parsecs. By 'direct' we mean that we exclude methods like chemical evolution models, n-particle simulations, etc, the results of which strongly depend on a number of hypotheses. Among the direct methods we admit the determination of the pattern speed by integrating the orbits of open clusters back to their birthplaces. The result, $R_{CR}$ = (1.06 $\pm$ 0.08)\rsun, does not have any strong dependence on the rotation curve, as shown by \cite{Dias2005}. Note that a contestation of this result, claiming that different arms have different velocities, was refuted by \cite{Lepine11} (L+8 hereafter). Other direct observations are the position of the ring-shaped gas void at corotation (\citealt{Amores2009}), and the position of the square-shaped spiral arm associated with the 4:1 resonance (L+8). These observations of resonances point to a slightly larger $R_{CR}$, but still within the error bar of the determination of $R_{CR}$ mentioned above. We adopt here \rsun = 7.5 kpc and $R_{CR}$= 8.1 kpc. The choice of the short scale of the Galaxy instead of \rsun= 8.5 kpc recommended by the IAU is based on many recent measurements, as discussed in detail  by L+8. However, the conclusions reached in this paper do not depend on that choice.

We illustrate in Figure \ref{fig1} the behaviour of two different tracers of chemical abundance in the region of corotation of our Galaxy. In Panel \ref{fig1}a we show the Fe abundance of open clusters as a function of radius, taken from L+8. There is a clear step down in metallicity and a gap in the distribution of objects at 8.5 kpc. This step was first discovered by \cite{Twarog1997}, but its connection with corotation was not noticed at that time. The distributions of metallicity shown in the figure are satisfactorily described by two plateaux, indicated by the vertical dashed lines. There is possibly some overlap in galactic radius between the high level and low level metallicity plateaux. Whereas this is one of the clearest evidences for the existence of a metallicity discontinuity, it is not the best choice for a comparison with other galaxies. In extragalactic studies the focus is on O abundance in \hii regions, while the Fe gradients are not available. Furthermore, Open Clusters are objects older than \hii regions, and their age give them the possibility of migrating out from their birthplace radius, as discussed by L+8. For instance the clusters situated beyond 13 kpc are old clusters that were probably born at a smaller galactic radius. Therefore, the extension of the low-level plateau to large radii probably does not correspond to a real metallicity plateau of the local gas. Similarly, the large radial overlap between the high-level and low-level plateaux could be explained by the radial migration of the clusters.

In Panel \ref{fig1}b we show the oxygen abundance data of OB stars
of \cite*{Daflon2004}. This is the best example of Galactic
abundance gradient to be compared with those of other galaxies,
since the objects are quite young (less than a few Myrs, which
excludes the possibility of migration) and oxygen is the element
being measured. Two straight lines fit well the data, and the step
of the gradient is quite clear. Beyond the corotation, the gradient
presents a positive slope. As it will be shown in later sections,
such positive slopes are not rare in other galaxies as well.

In the two panels of Figure \ref{fig1} we also fitted the data with polynomials of different degrees (we selected the degree which gives the best fit, according to the discussion in Section 4). It is remarkable that the fitted polynomials show a point of inflexion that coincides with $R_{CR}$. This analysis suggests that when a break or step of the gradient is clearly visible, the best approach is to adjust two independent lines, one on each side of this transition point. However, when the data presents only smooth variations of the slope, the minima of a fitted polynomial can be a good indicator of the position of corotation.

We should mention that there are in the literature on our Galaxy many works presenting the abundance of some elements as a function of radius. The results are based on all sorts of methods and objects. Only rarely one can clearly distinguish the discontinuities in the metallicity at $R_{CR}$. The main reasons are either too few data points, large error bars in the distances or in the metallicities, or the use of old objects which had time to migrate.

\section{Data Sample and Normalization}

\par The data used in this work were compiled from the literature. We selected galaxies for which there were one or more papers giving their corotation radii, radial metallicity distributions and rotation curves. This criterion was fulfilled by 27 galaxies with their data distributed in 130 references. Table \ref{tbl-1} contains the basic data for these galaxies; the references are given in Tables \ref{tbl-a1} and \ref{tbl-a2} in the Appendix.

\begin{table*}
 \caption{Selected galaxies and their projection parameters (see details in the text). The symbols used are i for inclination and PA for the position angle, both in degrees, D is the distance to the objects, in Mpc,  and s is the scale (kpc/arcmin) for the deprojected galaxy. The indices between brackets are the references for the distance (D) and the rotation curve (V(r)) used along the paper.}
 \label{tbl-1}
 \centering
\begin{tabular}{lrrrrrrrrr}
 \hline
  \hline

\multicolumn{ 1}{l}{{\bf Galaxy}} & \multicolumn{ 1}{c}{{\bf AR}} & \multicolumn{ 1}{c}{{\bf Dec}} &  \multicolumn{ 1}{c}{{\bf i}} & \multicolumn{ 1}{c}{{\bf PA}} &  \multicolumn{ 1}{c}{{\bf D}} &  \multicolumn{ 1}{c}{{\bf s}} & \multicolumn{ 1}{c}{{\bf Method}} & \multicolumn{ 2}{c}{{\bf References}} \\
\multicolumn{ 1}{c}{{\bf }} & \multicolumn{ 1}{c}{{\bf J2000}} & \multicolumn{ 1}{c}{{\bf J2000}} & \multicolumn{ 1}{c}{{\bf [\dg]}} & {\it {\bf (J2000) [\dg]}} & {\it {\bf [Mpc]}} & {\it {\bf [kpc/\pr]}} &   {\it {\bf }} &  {\it {\bf D}} & {\it {\bf V(r)}} \\
\hline
        IC0342 &  03:46:48.5 &  +68:05:46 &    21 &    130 &    3.3 &   0.96 &   Cph & [100] &     [113] \\
 NGC0224 (M31) &  00:42:44.3 &  +41:16:09 &    74 &     36 &   0.73 &   0.21 &   Cph & [126] &     [113] \\
 NGC0598 (M33) &  01:33:50.9 &  +30:39:36 &    53 &     25 &    0.8 &   0.23 &   Cph &  [67] &     [113] \\
 NGC0628 (M74) &  01:36:41.8 &  +15:47:00 &    16 &     25 &    6.7 &   1.95 &  SNII & [127] & [23]+[58] \\
 NGC1068 (M77) &  02:42:40.7 &  -00:00:48 &    32 &     72 &   15.7 &   4.57 &   Hub &  [24] &     [113] \\
       NGC1232 &  03:09:45.5 &  -20:34:46 &    20 &    269 &     21 &   6.11 &   Hub &  [24] &     [120] \\
       NGC1365 &  03:33:36.4 &  -36:08:25 &    46 &    222 &   18.3 &   5.32 &   Cph & [111] &     [113] \\
       NGC1530 &  04:23:27.1 &  +75:17:44 &    50 &     30 &   35.8 &  10.41 &   Hub &  [24] &      [72] \\
       NGC1566 &  04:20:00.4 &  -54:56:16 &    16 &     60 &   18.2 &   5.29 &   Hub &  [24] &      [86] \\
       NGC2403 &  07:36:51.4 &  +65:36:09 &    60 &    123 &   3.47 &   1.01 &  SNII & [128] &     [113] \\
       NGC2543 &  08:12:57.9 &  +36:15:17 &    60 &     46 &   33.5 &   9.74 &   Hub &  [24] &      [72] \\
       NGC2903 &  09:32:10.1 &  +21:30:03 &    69 &     21 &   6.52 &    1.9 &   Hub &  [24] &     [113] \\
 NGC3031 (M81) &  09:55:33.2 &  +69:03:55 &    60 &    157 &   3.63 &   1.06 &   Cph &  [41] &     [113] \\
       NGC3319 &  10:39:09.4 &  +41:41:12 &    58 &     42 &   14.3 &   4.16 &   Cph & [101] & [79]+[48] \\
       NGC3359 &  10:46:36.8 &  +63:13:27 &    51 &    172 &   14.4 &   4.19 &   Hub & [134] &       [3] \\
       NGC3621 &  11:18:16.5 &  -32:48:51 &    65 &    345 &    6.6 &   1.92 &   Cph &  [42] &       [7] \\
 NGC4254 (M99) &  12:18:49.6 &  +14:24:59 &    41 &     68 &   32.3 &    9.4 &   Hub &  [24] &      [64] \\
NGC4321 (M100) &  12:22:54.9 &  +15:49:21 &    27 &    146 &   16.1 &   4.68 &   Cph &  [39] &     [113] \\
 NGC4736 (M94) &  12:50:53.0 &  +41:07:14 &    50 &    125 &   4.93 &   1.43 &   Hub &  [24] &     [113] \\
       NGC4826 &  12:56:43.7 &  +21:40:58 &    59 &    114 &   5.48 &   1.59 &   Hub &  [24] &      [94] \\
       NGC5033 &  13:13:27.5 &  +36:35:38 &    63 &    170 &   12.7 &   3.69 &   Hub &  [24] &     [113] \\
 NGC5055 (M63) &  13:15:49.3 &  +42:01:45 &    57 &    107 &   15.7 &   4.57 &  SNIa & [107] &     [113] \\
 NGC5194 (M51) &  13:29:52.7 &  +47:11:43 &    20 &    170 &    8.9 &   2.59 &  SNII &  [26] &     [113] \\
 NGC5236 (M83) &  13:37:00.9 &  -29:51:56 &    24 &     45 &    4.5 &   1.31 &   Cph & [117] &     [113] \\
NGC5457 (M101) &  14:03:12.6 &  +54:20:57 &    18 &     39 &    7.4 &   2.15 &   Cph &  [57] &     [113] \\
       NGC6946 &  20:34:52.3 &  +60:09:14 &    44 &     58 &   3.76 &   1.09 &   Hub &  [24] &     [113] \\
       NGC7479 &  23:04:56.6 &  +12:19:22 &    46 &     41 &   34.9 &  10.15 &   Hub &  [24] &      [49] \\
 \hline
  \hline
 \end{tabular}

\end{table*}

As expected, the measurements made by many groups of research and at different epochs are based on a variety of methods and assumptions.  For this reason the following criteria were used to select the data and to standardize the measurements:

\begin{description}

\item \noindent {\it Geometrical Parameters}: The distances of objects to the centre of a galaxy depend fundamentally on the distance to the galaxy and on projections parameters of its disk on the plane of the sky (inclination and position angle). The first step of this work was to recalculate the radial distances of the objects given in different papers, adopting for each galaxy the distance and projection angles that we determined. We compared the results obtained by the traditional method of fitting ellipses to isophotal curves, and from spiral arms fitting \citep{Ma01} in images available in DSS. For the present sample of galaxies the results are in agreement with each other and also with the results on the literature, especially those derived from velocity fields (see references in the appendix). Since the differences of the resulting angles based on the different methods are smaller than $15^\circ$, we adopted the median of all results to derive the values in Table \ref{tbl-1}. The most critical discrepancies (of almost a factor 2 in some cases) were on the distances to the galaxies. We selected the distances given in the literature according the method used to determine them. The preferences among the methods follows the sequence: 1-) Cepheids; 2-) Supernovae; 3-) Hubble Flow, adopting H=73 km/s/Mpc.

\item
\item \noindent {\it Metallicity Distributions}: The most common way to measure the metallicity distribution in spiral galaxies makes use of the nebular lines from \hii regions. In a strict sense, "metallicity" refers to the abundance of elements heavier than Helium. Among these elements oxygen is the most abundant and easiest to be measured. It is also the element which has abundances available for the largest number of galaxies. Consequently, we concentrated our study on this element. Since O is mainly released by Type II Supernovae, objects that are strongly associated with the spiral arms of galaxies, it is also a convenient element to investigate the effects of the corotation resonance on the radial metallicity distributions. There are several calibrators that allows calculate the oxygen abundances based on the ratio of line fluxes, going from the most direct methods, based on temperature-sensitive line ratios (Te-method hereafter), to  some sort of statistical methods that do no use oxygen line fluxes. Since each calibrator may be affected by different systematic effects (see \cite{Pilyugin03} or \cite{Kewley08} for example). For this reason we propose to study the changes in the radial metallicity distributions of a galaxy using all calibrators available and verifying the consistence between the results.
\item
\item \noindent {\it Rotation Curves}: It is not uncommon to find differences on the rotation curve of a galaxy in different papers, especially when they were observed with different techniques. Preference was given to:  1-) Rotation curves composed by radio and optical observations; 2-) when optical and radio observations are available, but they have been presented separately, we joined them to use the resulting curve; 3-) rotation curves from radio velocity fields only; 4-) rotation curves from optical velocity fields only; 5-) rotation curves from long slit data. All rotation curves were recalculated considering the distances, inclinations and position angles we adopt in this work (Table \ref{tbl-1}).
\item
\item \noindent {\it Corotation radius and/or spiral pattern speed}: These two parameters are easily converted from one to the other if we know the rotation curve ($\omega_p = V(R_{CR})/R_{CR}$). In principle we did not make any distinction between methods to evaluate the corotation radii or the spiral pattern speeds. However, we confront the values obtained in the literature with the constraints imposed by the rotation curve and the extension of the spiral structure. When several measurements where available in the literature, we only considered the median of the values which are not in conflict with these constraints. In our effort to standardize the radii of corotation we faced two situations: 1-) cases when there are differences between our standard distance to the galaxy and the distances adopted by the consulted references, but similar rotation curves were considered; 2-) cases where there is a significant difference between our standard rotation curve and the rotation curve from the references on corotation. In the first case we just scaled the corotation radius or the spiral pattern speed to the new distance. In the second case it is necessary to convert the pattern speed of the original reference to corotation radius, using the original rotation curve and then scale this radius to the adopted distance. The final corrected pattern speed is obtained by converting back the scaled corotation radius using the standard rotation curve. The uncertainty limits (or extremities of the error bars) of the original paper are converted to the new system using the same procedure. Since rotation curves are usually not linear, this conversion produces some asymmetric error bars. In these cases we adopted the largest values of uncertainties.
\end{description}

\section{Methods to Identify Breaks in the Metallicity Distribution}
\subsection{Preliminary tests with Monte-Carlo simulations}

We have seen from the example of our Galaxy, that in some cases, when there are plateaux separated by a step, one before and one after the corotation radius, an inflexion point reveals the radius of the break. However, we also found cases of galaxies for which a change in the gradient slope is clearly associated with corotation, without a noticeable step in the level of metallicity. In such cases a good representation of the data can be obtained by fitting them with two lines connected at their intersection. There are still other cases where there is a minimum in the metallicity at the corotation radius, because the slope of the gradient changes from negative to positive. In such cases a polynomial fitting is a convenient procedure to identify the minimum.

A question that arises is: how can we check that the function that we adopt to represent the abundance distribution is a better choice than a straight line, which is the usually accepted fit?

In order to establish the significance of a model with a break in the gradient, \cite{Zahid2011} has performed likelihood ratio F-test to the galaxies they studied, but we did not adopt the same approach. In many cases it is not possible to standardize the data in the literature in reason of the lack of information on the flux of some strategic lines. Another problem arises mainly from the different ways the uncertainties are evaluated in the literature, and the F-test is particularly sensible to these differences. This problem would force us to discard many objects in a sample that it is already small, so we opted to follow an intermediary procedure. We compared the chi-square deviations of the data from the fitted curve, divided by the number of degrees of freedom (hereafter $\chi^2$/dof). However, even when this parameter is smaller for a model with a break than for the straight line, there is still a probability of the straight line being the "real" one. We must then be able to attribute relative probabilities to the different functions that can be fitted. We performed Monte-Carlo simulations to answer this question.

In the tests we made, a random noise was introduced on the top of a well behaved law, that we will call the "real" law, and then attempted to recover this original law by an automatic fitting procedure. For instance, it may happen that even starting from a real straight line, the fitting procedure tells us that a third order polynomial is the best fit, based on the chi-square test. By repeating the experiment hundreds of times, it is possible to infer the probability of such an event to occur.

The metallicity distributions were simulated starting with a
straight line which represents the average oxygen abundance for
spiral galaxies derived from the tables 1 and 3 of \cite{Zaritsky94}
(Eq. \ref{eq1}).

\bigskip
\begin{equation}
Z(r)=-0.09 \cdot r+ 8.98 \label{eq1}
\end{equation}
\bigskip

In a first part of the tests we generated data points with a random dispersion of 0.3 dex relative to this line. This dispersion is typical of the observed data. In a second part we created a new "real" radial metallicity distributions with a minimum with a depth of 0.3 dex with respect to (Eq. \ref{eq1}), by adding a cosine shape function in the middle of the distribution. The width of the minimum is similar to the typical uncertainty for the corotation radii (3 kpc). After that we generated data points with 0.3 dex dispersion with respect to the new distribution. The experiment was repeated using two connected lines changing the slope in the middle of the distribution, instead of a minimum, as being the original function. The change of slope was taken to occur at 0.3 dex below the straight line.

\begin{figure}
\includegraphics[width=84mm]{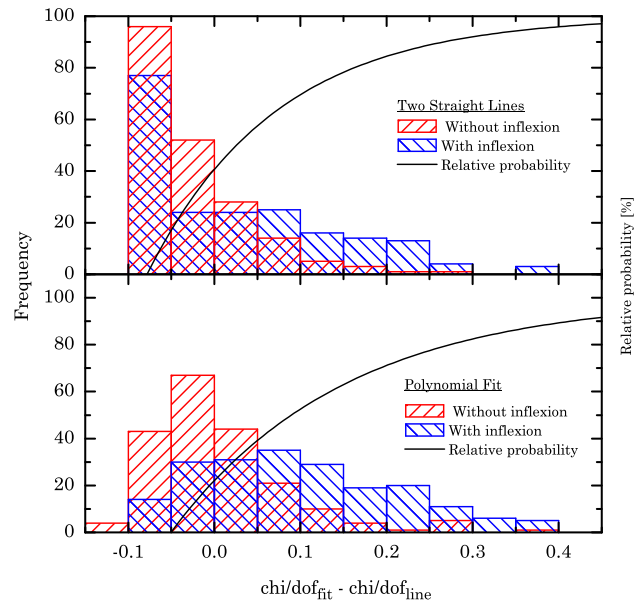}
\caption{Results of Monte-Carlo simulations. A random distribution of points is generated around an original function (two straight lines for the upper panel, a function with a minimum for the lower panel and a straight line for the two panels). The original function and a straight line are fit to both set of data in each panel (with and without inflexions), giving us $\chi^2$/dof(fit) and $\chi^2$/dof(line) for each case. The histograms compare the frequency of the cases for which the correct function is recovered (represented with dropping lines hatch pattern) and the frequency of the straight line being the best result (rising lines hatch pattern), both as a function of $\chi^2$/dof(fit)-$\chi^2$/dof(line). Each histogram is the result of 200 simulations. The continuous line represents the probability of the correct function to be recovered, which is read in percentage in the scale.}
 \label{fig3}
\end{figure}

For each original or real function we performed 200 simulations by generating different sets of data points, and analyzed these sets as follows. The simulated data were analyzed by automatically fitting different functions: a linear fit, a 4th order polynomial, or two connected straight lines with different slopes. For each fit we recorded the differences between the $\chi^2$/dof of the fitted models and that of the single straight line fit. The probability of having a correct identification as a function of the difference of $\chi^2$/dof are presented in Figure \ref{fig3}. Let us call $\alpha$ the parameter [$\chi^2$/dof(model)- $\chi^2$/dof(straight line)]

As it could be expected there is a range of the parameter $\alpha$, corresponding to small differences of the chi-squares of the two fits, for which the quality of the fits is similar, and for which the probabilities of one or the other functions being the real one are similar. However, when $\alpha$ becomes larger than 0.025, the initial model function becomes the most probable one, compared to a straight line. For this reason, we will classify as "strong point" the radial metallicity distribution of galaxies for which $\alpha >$ 0.025.

Interestingly, we found that even in situations where the above argument based on the value of $\alpha$  does not allow us to eliminate the possibility of a straight line, in most cases the polynomial fit provides a quite good result, in the sense that it recovers the correct position of the minimum, when the original function had one.

\subsection{Adopted procedure}

The radial abundance distribution data of each galaxy were treated by a fully numerical and automatic analysis procedure, as follows:

\begin{enumerate}
\renewcommand{\theenumi}{(\arabic{enumi})}

\item \noindent The observed metallicity distribution was fitted by single straight line, two connected straight lines (4 degrees of freedom), two straight lines with a step (5 dregrees of freedom), and then 4 polynomials, of order 2 to 5 (the number of degrees of freedom varies from 3 to 6).  We verified that there is no need to go to higher orders. The uncertainties were taken to be equal for all the data points.

\item \noindent For each fit the $\chi^2$/dof was evaluated according to \cite{numrec1992}. A description of a similar application of this recipe can be seen in \citet{Scarano11};

\item \noindent For the two connected straight lines and the two straight lines with a step, the point of connection or the step position respectively was assumed to be the point where the gradient break occurs, included in Table \ref{tbl-2}. For the polynomials we identified the minima and inflexions by studying the derivatives of the fitted curves. If the step function fitted the data better than the other fits, the inflexions points of different polynomials were used to determine the error on the radius of the metallicity break. If the two connected straight lines fitted the data better than the other fits, the polynomial minima were used to calculate the error on the radius of the metallicity break;

\item \noindent Taking into account only the results for which the $\chi^2$/dof was lower than that obtained with the linear fit, we registered the radius of the minimum for each function fitted. If a fitted polynomial presented more than one minimum, we only considered the one situated near the minimum of the other polynomials. With these values we estimated the position of the break in the gradient and its dispersion. We adopted the weighted mean of the positions, using as weights the inverse of the $\chi^2$/dof.

\end{enumerate}

\begin{figure*}
\includegraphics[width=180mm]{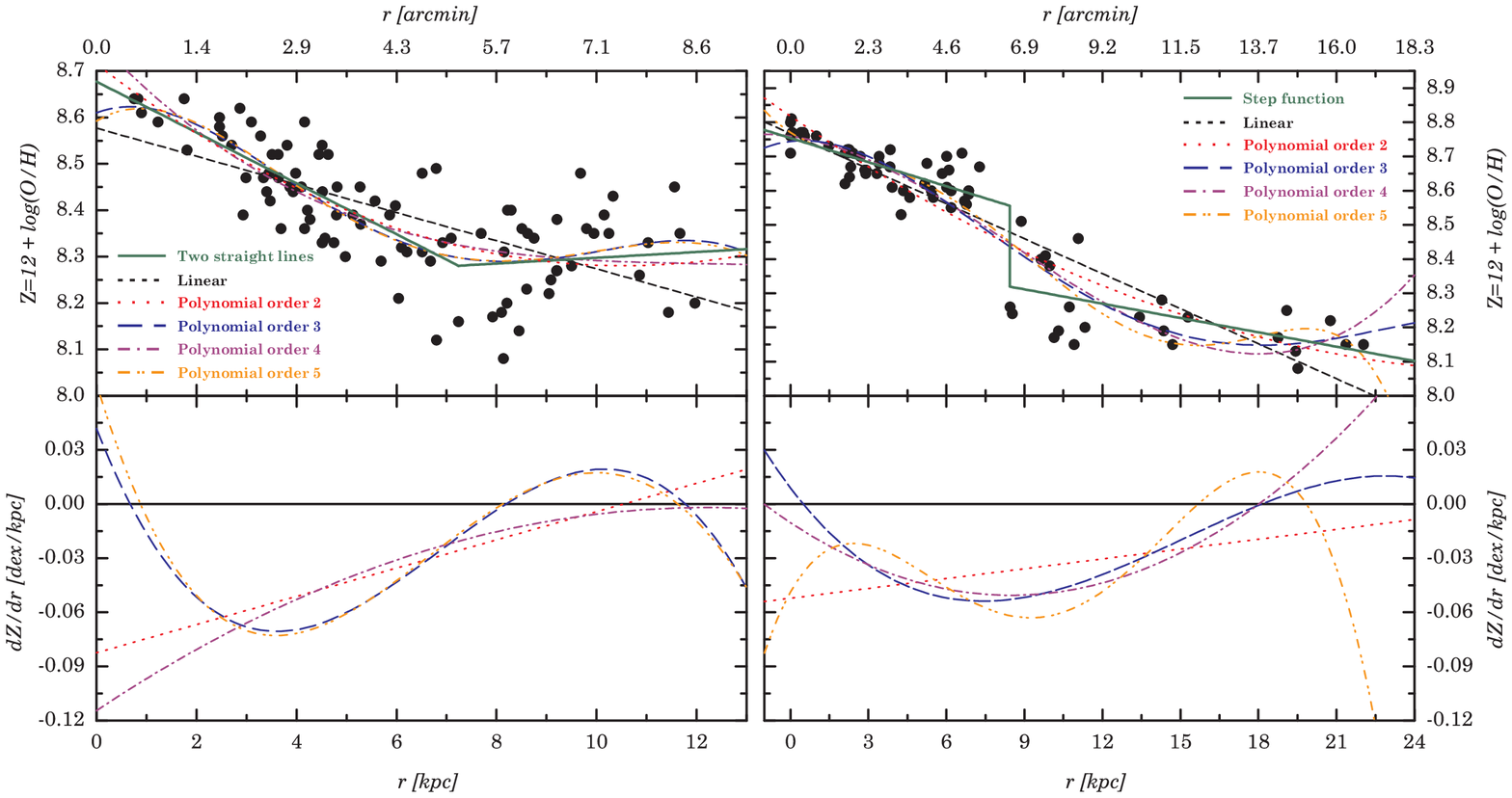} \caption{Radial oxygen distribution (top panel) and metallicity gradients (bottom panel) for the galaxies NGC3359 (left) and NGC5236 (right). The models fitted to the data are represented by different line types, as indicated in the legend.}
 \label{fig2}
\end{figure*}

Figure \ref{fig2} illustrates the application of this procedure to the galaxies NGC3359 and NGC5236. These galaxies are representative of the two classes of breaks found in the literature: smooth breaks and step breaks. For NGC3359 we can see the general behaviour of the first class of objects: the radial metallicity distribution changes smoothly, and it is better fitted by two connected straight lines. Higher order polynomials also produce good fits, with minima close to the intersection of the two connected straight lines. Lower order polynomials also tend to reveal breaks, but depending on the data dispersion and the steepness of the slopes, the minima found are farther than those obtained with other models. On the other hand NGC5236 represents the second class of breaks. In general, for these cases, only the step function fits the data properly. As it could be expected, the minima of polynomial fits are not correlated with the step position. However, the inflexions of the polynomials tend to be near the position where the gradients of abundances changes abruptly.

\section{Results}

\par We performed the analysis of the galaxies listed in Table \ref{tbl-1}, following the procedures described in the previous section. The data needed for this study were found in the literature, the references being given in Tables \ref{tbl-a1} and \ref{tbl-a2} (Appendix). The results are presented in Table \ref{tbl-2}, where the breaks or minima in the metallicity distribution obtained by the automatic fitting method are given.

\begin{table*}
 \caption{Corotation radii and breaks of metallicity gradients. The index n is used to represent the galaxy in the graphs. R$_{CR}$ and $\Omega_p$ are respectively the median of the corotation radii and the spiral pattern speeds found in the literature, discarding bad values, R$_{dZ}$ is the metallicity break obtained by the procedure written in the text, and $\Omega_{dZ}$ is the angular velocity of the rotation curve at R$_{dZ}$. The "Notes" column present comments on the quality of the data or of the fit to obtain the metallicity breaks (S= strong point, W= Weak point, FP= too few points in the metallicity distribution, see text for explanation of S and W). The last column gives the references for the radial metallicity distributions according to Table \ref{tbl-a1} in the Appendix.}

 \label{tbl-2}
 \centering
\begin{tabular}{llllllll}
 \hline
  \hline

 {\bf n} &    {\bf Galaxy} &   {\it {\bf R$_{CR}$}} &  {\it {\bf $\Omega_p$}} &   {\it {\bf R$_{dZ}$}} &  {\it {\bf $\Omega_{dZ}$}} &   {\it {\bf Notes}} &    {\it {\bf References}} \\
  {\bf } &          {\bf } &        {\it {\bf [kpc]}} &   {\it {\bf [km/s/kpc]}} &        {\it {\bf [kpc]}} &    {\it {\bf [km/s/kpc]}} &       {\it {\bf }} &         {\it {\bf }} \\
  \hline
    \hline
 1 &    IC0342 &    10.2 $\pm$ 3.1 &    19.0 $\pm$ 7.4 &                 - &                 - &    FP &                                  [76], [89], [125]  \\
   &           &                   &                   &                   &                   &       &                                                     \\
 2 &   NGC0224 &    15.6 $\pm$ 1.1 &    18.1 $\pm$ 1.1 &    15.3 $\pm$ 3.6 &    18.2 $\pm$ 4.6 &    W &              [10], [62], [89], [125], [133], [135]  \\
   &           &                   &                   &                   &                   &       &                                                     \\
 3 &   NGC0598 &     4.3 $\pm$ 1.4 &    24.0 $\pm$ 6.1 &     5.4 $\pm$ 1.3 &    19.1 $\pm$ 4.4 &    W &         [10], [22], [54], [62], [71], [76], [89],   \\
   &           &                   &                   &                   &                   &       &                                [109], [125], [135]  \\
 4 &   NGC0628 &     4.6 $\pm$ 1.2 &    19.2 $\pm$ 7.6 &     4.8 $\pm$ 1.8 &    18.4 $\pm$ 6.5 &    W &          [5], [10], [17], [38], [76], [80], [88],   \\
   &           &                   &                   &                   &                   &       &    [89], [109], [121], [122], [125], [135], [137]  \\
 5 &   NGC1068 &     3.2 $\pm$ 0.1 &    67.5 $\pm$ 2.1 &                 - &                 - &    FP &             [36], [83], [89], [121], [122], [135]   \\
   &           &                   &                   &                   &                   &       &                                                     \\
 6 &   NGC1232 &    12.6 $\pm$ 3.0 &    25.6 $\pm$ 6.1 &    15.0 $\pm$ 2.1 &    21.6 $\pm$ 3.1 &   S &                [9], [17], [88], [89], [121], [122]  \\
   &           &                   &                   &                   &                   &       &                                                     \\
 7 &   NGC1365 &    20.7 $\pm$ 4.0 &    11.5 $\pm$ 5.6 &    20.4 $\pm$ 5.3 &    11.6 $\pm$ 3.8 &    S &    [9], [28], [62], [87], [89], [98], [125], [135]  \\
   &           &                   &                   &                   &                   &       &                                                     \\
 8 &   NGC1530 &    13.4 $\pm$ 5.0 &     8.2 $\pm$ 5.0 &                 - &                 - &    FP &                                               [72]  \\
   &           &                   &                   &                   &                   &       &                                                     \\
 9 &   NGC1566 &     8.8 $\pm$ 1.3 &    23.6 $\pm$ 5.0 &                 - &                 - &    FP &                          [62], [109], [125], [135]  \\
   &           &                   &                   &                   &                   &       &                                                     \\
10 &   NGC2403 &     5.1 $\pm$ 3.0 &       25 $\pm$ 15 &     6.6 $\pm$ 0.6 &    19.2 $\pm$ 1.7 &    W &               [10], [40], [47], [76], [80], [87],   \\
   &           &                   &                   &                   &                   &       &            [89], [109], [121], [122], [125], [135]  \\
11 &   NGC2543 &     6.0 $\pm$ 2.2 &    25.9 $\pm$ 9.3 &                 - &                 - &    FP &                                               [72]  \\
   &           &                   &                   &                   &                   &       &                                                     \\
12 &   NGC2903 &     4.1 $\pm$ 0.1 &    69.9 $\pm$ 2.7 &     5.0 $\pm$ 0.8 &   56.4 $\pm$ 10.5 &    S &      [62], [76], [88], [89], [109], [121], [122],   \\
   &           &                   &                   &                   &                   &       &                                       [125], [135]  \\
13 &   NGC3031 &    11.0 $\pm$ 2.0 &    18.2 $\pm$ 5.0 &     9.8 $\pm$ 3.4 &   21.7 $\pm$ 12.1 &    S &  [10], [54], [62], [85], [89], [114], [125], [135]  \\
   &           &                   &                   &                   &                   &       &                                                     \\
14 &   NGC3319 &     7.4 $\pm$ 2.0 &     9.8 $\pm$ 0.7 &     6.6 $\pm$ 0.7 &    10.1 $\pm$ 0.3 &    W &                                        [62], [135]  \\
   &           &                   &                   &                   &                   &       &                                                     \\
15 &   NGC3359 &     6.1 $\pm$ 0.6 &    20.7 $\pm$ 2.7 &     7.9 $\pm$ 0.5 &    16.2 $\pm$ 0.8 &    S &                                        [73], [134]  \\
   &           &                   &                   &                   &                   &       &                                                     \\
16 &   NGC3621 &     6.2 $\pm$ 2.0 &    22.0 $\pm$ 4.7 &     8.7 $\pm$ 1.0 &    16.1 $\pm$ 1.8 &    S &                                               [12]  \\
   &           &                   &                   &                   &                   &       &                                                     \\
17 &   NGC4254 &    13.6 $\pm$ 2.2 &    10.5 $\pm$ 1.6 &    11.5 $\pm$ 3.0 &    12.6 $\pm$ 2.2 &    W &       [62], [76], [80], [88], [89], [109], [112],   \\
   &           &                   &                   &                   &                   &       &                                       [125], [135]  \\
18 &   NGC4321 &    10.5 $\pm$ 1.3 &    25.8 $\pm$ 3.0 &                 - &                 - &    FP &        [28], [62], [76], [80], [88], [89], [109],   \\
   &           &                   &                   &                   &                   &       &                                [112], [125], [135]  \\
19 &   NGC4736 &     3.1 $\pm$ 0.4 &    56.0 $\pm$ 5.0 &                 - &                 - &    FP &   [10], [62], [75], [76], [83], [89], [125], [135]  \\
   &           &                   &                   &                   &                   &       &                                                     \\
20 &   NGC4826 &     1.2 $\pm$ 0.1 &    56.2 $\pm$ 5.9 &                 - &                 - &    FP &                                               [83]  \\
   &           &                   &                   &                   &                   &       &                                                     \\
21 &   NGC5033 &    10.4 $\pm$ 0.7 &    25.1 $\pm$ 1.0 &                 - &                 - &    FP &                            [62], [80], [89], [135]  \\
   &           &                   &                   &                   &                   &       &                                                     \\
22 &   NGC5055 &    14.7 $\pm$ 4.4 &    14.6 $\pm$ 3.8 &                 - &                 - &    FP &                     [62], [76], [89], [125], [135]  \\
   &           &                   &                   &                   &                   &       &                                                     \\
23 &   NGC5194 &     5.6 $\pm$ 0.6 &    45.6 $\pm$ 4.3 &     6.8 $\pm$ 0.5 &    37.0 $\pm$ 2.7 &    S &          [8], [10], [62], [76], [80], [88], [89],   \\
   &           &                   &                   &                   &                   &       &                                       [125], [135]  \\
24 &   NGC5236 &     8.1 $\pm$ 2.0 &    22.5 $\pm$ 6.5 &     8.4 $\pm$ 3.0 &    21.8 $\pm$ 9.9 &    S &    [9], [10], [11], [62], [88], [89], [125], [135]  \\
   &           &                   &                   &                   &                   &       &                                                     \\
25 &   NGC5457 &    16.7 $\pm$ 2.4 &     5.5 $\pm$ 1.2 &    17.7 $\pm$ 2.2 &     5.9 $\pm$ 0.5 &    S &         [28], [54], [61], [76], [87], [88], [89],   \\
   &           &                   &                   &                   &                   &       &                  [109], [121], [122], [125], [135]  \\
26 &   NGC6946 &     4.2 $\pm$ 0.4 &    53.8 $\pm$ 1.9 &                 - &                 - &    FP &         [5], [38], [62], [76], [88], [89], [109],   \\
   &           &                   &                   &                   &                   &       &                                       [125], [135]  \\
27 &   NGC7479 &    14.6 $\pm$ 2.7 &    12.8 $\pm$ 2.4 &    16.6 $\pm$ 0.6 &    11.0 $\pm$ 0.5 &    S &                                   [72], [74], [89]  \\
   &           &                   &                   &                   &                   &       &                                                     \\
 \hline
  \hline

\end{tabular}

\end{table*}

\par
\begin{figure*}
\includegraphics{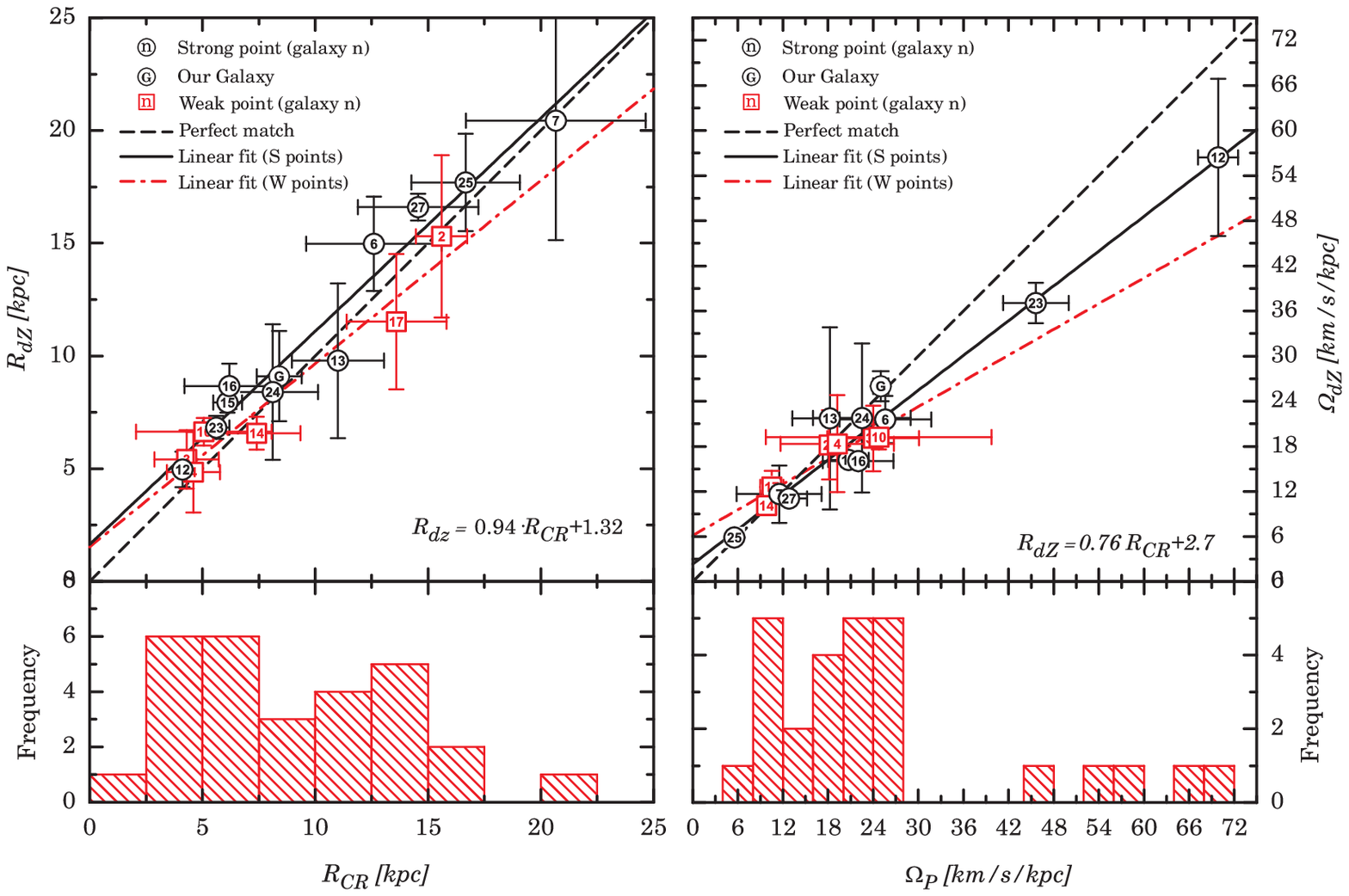} \caption{Correlations between the corotation radii and the metallicity breaks radii (left panel) and the spiral pattern speed and the angular velocity expected at the metallicity breaks radii (right panel). We labelled the data points according the index sequence of galaxies in Table \ref{tbl-2} ("n" column). Our Galaxy is represented  by a "G" letter; the dashed line corresponds to the perfect match line, the dashed-dotted line is the linear fit considering all dataset; the continuous lines take into account only galaxies with the strongest evidences of breaks in their oxygen distribution (see text). The histograms at the bottom part of each correlation graph show the number of objects per bin of data (corotation or spiral pattern speed), taking into account all galaxies presented in Table 3 (bin size equals the mean of uncertainties of these data).}
 \label{fig4}
\end{figure*}

The main result of the present work is the existence of a clear correlation between corotation radii and breaks in the metallicity distribution, shown in Figure \ref{fig4}, left panel. The linear fit to the data presented in this figure is very near the one-to-one match, which means that the two occurrences are spatially associated in the galaxies studied here. As expected from the inverse relation between corotation radius and pattern speed, a good correlation can also be seen in a plot of angular speeds (angular velocity of the rotation curve at the radius of metallicity break versus angular pattern speed), in the right panel of the figure. The uncertainty is large for some points in this case, among other reasons, due to the non-linearity in the process of conversion of the data from  corotation radius to angular velocity, especially when no uncertainties were mentioned in the literature and we had to assume the median uncertainty in the corotation radii of the other references (sigma $\sim 2$kpc).

In spite of our limited sample some side results can be seem at bottom of Figure \ref{fig4}. The distribution of pattern rotation speeds shows a broad peak at 24 km/s/kpc, which reveals that our Galaxy has a "typical" pattern angular velocity. This peak would be even more prominent if only the "strong" points were shown in the histogram. This result is certainly a clue to understand what determines the value of $\Omega_p$ in galaxies. On the other hand, the distribution of corotation radii (without any normalization to the size of the galaxy, in our work) is somewhat uniformly spread between about 3 kpc to 17 kpc. Considering that large spiral galaxies may reach typically a radius of 30 kpc \citep{Tully87} which should correspond to the outer limit for spiral arms (the Outer Lindblad Resonance), the cutoff seen in the corotation radius histogram of Figure \ref{fig4} is coherent with the expectation that the corotation radius must be near midway between the Lindblad resonances.

\section{Comments on results for individual galaxies}

We next present comments on the 27 galaxies included in Tables \ref{tbl-1} and \ref{tbl-2}, which are the galaxies for which we found determinations of both metallicity gradient and corotation radius. Some of these galaxies were not included in the correlation, for reasons that are explained in the notes, but they are included in our statistics on corotation radii. Of course, there are many more galaxies in the literature, not included in our study, for which only one of these two sets of data was found. One example is NGC1512, for which there is a nice set of measurements of O abundance by Bresolin et al. (2012), which shows the presence of a break in the slope of the distribution at 12-13 kpc. However, we did not find a determination of the corotation radius in the literature. We classified the galaxies that remained in the correlation shown in Figure \ref{fig4} in two classes, the "strong" and "weak" points. These denominations do not have any direct relation with the error bars which affect the points. We consider that we have a strong point when the following conditions were fulfilled: 1-) more than 16 measurements of O abundance at different radii are available, satisfying the condition prescribed by \cite{DutilRoy01}; 2-) at least two different calibration methods were used to determine the abundance and 3-) the breaks (minima, breaks or steps) in the abundance distribution derived from the different calibration methods are in agreement. Galaxies for which there are between 10 and 16 measurements at different radii or only studied using one single calibration method, even showing a clear break in the abundance distribution, are considered as weak points. A few galaxies which fulfil the "strong" conditions were converted to weak points for some other particularity like non-even distribution of the points, inhomogeneity of the metallicity sample, or other, as explained.

\begin{description}

\item \noindent {\it IC0342}: This galaxy was not included because only five \hii regions were observed.

\item
\item \noindent {\it NGC0224 (M31)}: It is difficult to see the spiral structure of this galaxy, due to its high inclination. There are many abundance measurements of HII regions. One set of data is that of \cite{Pilyugin04}. However, with these data, the best fit obtained is linear (without breaks in the slope). We attribute this to the scarcity of data beyond 10 kpc,  where the break occurs. A change of slope around 15 kpc can be seen in the data of \cite{Blair1982} and of \cite{Worthey+05}. The data of Worthey (34 objects) include, in addition to HII regions, stars observed with the Hubble Space Telescope, as distant as 50 kpc. The data are well fitted with two lines with a break at 15.2 kpc. The extended region with very little change in metallicity, from about 15 to 50 kpc, is well fitted by a straight line and makes the polynomial fits to look unsatisfactory. Although there is good agreement between the break and the corotation radius, we will consider this galaxy as a "weak" point, due to the inclusion of different types of objects in the determination of the gradient.

\item
\item \noindent {\it NGC0598 (M33)}: There are many discrepant determinations of corotation for this galaxy. \cite{Newton80} considers that the corotation radius can be 1.2 or 7.0 kpc (scaled to our system). \cite{Puerari93} determines a pattern speed which corresponds to corotation radius of 4.3 kpc. A recent work on the distribution of planetary nebula by \cite*{Magrini2009b} shows a gap at 3.8 kpc (also scaled), with a constant or slightly rising oxygen abundance beyond that radius. Larger (\cite{VilaCosta92} 5.9 kpc) and smaller (\cite*{Elmegreen92} 2.9 kpc) radii can be found as well in the literature. We adopted the median corotation radius of the data from the literature. This result is coherent with extension of the spiral structure. Concerning the metallicity distribution, recent results have been produced by \cite{Magrini2007} (28 points) and \cite{Bresolin2011} (46 points). In the work of Bresolin et al., there are some "best quality" points, but they do not reach the radius of corotation; the other points, collected from the literature, show a larger scattering and are best fitted by a single straight line. The data of Magrini et al. on planetary nebulae is best fitted by two lines with a break in the slope. This result is consistent with the break at the same radius, but considering that this result was obtained only for one calibration method, this galaxy is considered as a weak point.
\item

\item \noindent {\it NGC0628 (M74)}: {\bf The metallicity distribution of this galaxy was studied using two sets of data independently. One set is that of using the data by \cite{Rosales+11}, who obtained abundances of 96 regions using integral field spectroscopy with 4 different calibration methods. If we consider only the new data obtained by these authors, for two of these calibrations the fit of two lines with a step reveal a very smooth break in the metallicity gradient near 3.6 kpc. Better results are obtained when data from the literature are added to the new observations, as proposed by Rosales-Ortega et al. The best fitted model to all calibration methods present a step near 6.1 kpc. However, the histogram of radial breaks found as a function of the galactrocentric distance exhibit two peaks: one near 5 kpc and another near 8.5 kpc for all calibrations}. The other set of data that we analysed is that of \cite{Moustakas+10}, which present two distinct calibration methods, with 34 points. These are the theoretical calibration published by \citeauthor{KobulnickyKewley04} \citeyearpar[hereafter KK04]{KobulnickyKewley04} and the empirical calibration method by \citeauthor{PilyuginThuan05} \citeyearpar[hereafter PT05]{KobulnickyKewley04}. For the first calibration the data is best fitted with two straight lines or two lines with a step, with the breaks at 6.3 and 6.9 kpc respectively. The results of the PT05 calibration provide better fits than those of the other calibration, and are best adjusted with polynomials of order 4 and 5, followed by 2 lines, and two lines with a step. {\bf The average for this calibration, weighted by $\chi^2$, is a break at 4.8 kpc. This radius agrees with the corotation determination of \cite{Cepa90}, \cite{Elmegreen92} and \cite{VilaCosta92} which give an average of 4.6 kpc. Interestingly,  for the data of this calibration the fit of two lines with a step indicates the existence of a step at about 7 kpc. This result is not much different from the second peak in the histogram that we mentioned for the data analyzed by \cite{Rosales+11}. Taking into account the rotation curves provided by \cite{Kamphuis92} and \cite{Daigle06} we see that the corotation at 4.6 kpc provides a spiral pattern speed of 19.2 km/s/kpc. In this case the outer Lindblad resonance (the limit where the spiral arm can exist) would be placed at 7.2 kpc, very near the end of the visible spiral arms. Since beyond the end of the spiral structure there would not be an efficient and continuous process to enrich the medium, a break in metallicity is also expected at the outer Lindblad resonance, and we could attribute to this resonance the outer break seen in the histogram of radial breaks. However we do not exclude the possibility of corotation being between 7.5 and 8.5 kpc, since this value would be more consistent with the diffuse extension of the spiral structure seen in the \hi line \citep{Walter2008}. Also we cannot discard the superposition of multiple pattern speeds.} Due to the lack of close agreement between the  calibration methods, we classify this galaxy as a weak point.
\item

\item \noindent {\it NGC1068}: Only 9 \hii regions have been measured, and in addition, they are poorly distributed along the disk. We exclude this galaxy from our correlation analysis.
\item

\item \noindent {\it NGC1232}: Galaxy sampled by 16 \hii regions by \cite{Pilyugin04}, originally from \cite{vanZee98}, to which new sources were added by \cite{Bresolin+05}. Since there are points in common in the two samples, one can see that the two calibrations are in good agreement, so that we can add the samples, resulting in a total of 23 \hii regions. The data is well fitted by polynomials of order 3,4 and 5, and by two lines. The radius of the break from  these different fits were averaged. This galaxy can be considered as a  strong point; the coincidence of the break with the corotation radius is satisfactory.
\item

\item \noindent {\it NGC1365}: Pilyugin collected about 70 points in the literature for NGC1365, most of which (55) are due to \cite{RoyWalsh97}, who were the first to detect the break in the metallicity slope of this galaxy. Later \cite{Pilyugin03} put in doubt the  reality of the break, which could be an artefact due to the existence of two branches in the calibration of the R23 method.  We have several arguments telling us that the break is real: 1-) Roy and Walsh compared 3 different methods of calibration for his data, and not all of them necessarily affected by the same systematic error; 2-) \cite{Bresolin+05} added new data points which allow a comparison of the P-method with the Te-method. The two methods give the same values of 12+ log[O/H],  between 8.6 to 8.4 dex, in the range of distances from 15 to 20 kpc.  This suggests that the abundances found beyond 20 kpc, which are in the same abundance range, were not  affected by a systematic error in the Pilyugin's data; 3-) In the same paper in which the reality of the break is discussed, it is said that problem of calibration could affect abundances smaller than 8.2 dex. However, almost all the points are above this limit, in the case of NGC1365, even taking in to account the data dispersion of 0.2 dex, so there is no reason to believe that all data points could be affected by calibration errors. Consequently, we will use not only the data of  Pilyugin, presented in his Figure 2b, for our analysis, but also all calibrations presented by \cite{RoyWalsh97}. One difficulty that one has to face is that there are two solutions. The data is well fitted by polynomials of 3rd, 4th and 5th order, and by 2 lines, which give a minimum at about 20 kpc. This is the minimum which is shown with 2 lines in Pilyugin's Figure 2b. However, the data is also marginally well fitted with two lines with a step at 16.6 kpc, close to inflexion points revealed by the same 3rd, 4th and  5th order polynomials. It should be remembered that in the presence of a step, we must look for the inflexion points of the polynomials, and not for the minima, as explained in Section 2. There seems to be a robust case for the presence of the step, more clearly seen in the data based on R23. Concerning corotation, there are several measurements in the literature, the median being 13.9 kpc, but  most of these values are incompatible with the extension of the spiral structure if we take into account the rotation curve by \cite{Sofue96}. The only value compatible with the observed spiral arms is an alternetive value mentioned by \cite{Vera01} of 20.7 kpc (re-scaled).  A recent detailed study of NGC1365 By \cite{Elmegreen2009} is of particular interest, since they detected a radial flow of gas along the spiral arms, in opposite directions on the two sides of corotation. The large error bar attributed to the radius of the break accounts for the two solutions (about 16 kpc and about 20 kpc).

\item
\item \noindent {\it NGC1530}:  Its metallicity distribution has only three observed \hii regions, so that it was discarded from our  analysis.
\item

\item \noindent {\it NGC1566}: Grand-design galaxy with six \hii regions observed, so that it was discarded.
\item

\item \noindent {\it NGC2403}: Flocculent galaxy with 45 \hii regions compiled by \cite{Moustakas+10}. For one of their calibrations (KK04 method) all models proposed here fit the data better than a straight line and show the presence of a break, while for their other calibration (PT05 method) the data is best fitted by a straight line. For this reason we classify it as a weak point. There is only one reference giving the corotation radius of this galaxy \citep{VilaCosta92}, which is compatible with the radius of the break in metallicity that we found.
\item

\item \noindent {\it NGC2543}: Galaxy with too few observed \hii regions. Although it was discarded from our analysis, it is interesting to mention that at least visually there is a step in the metallicity distribution of this galaxy at the same position where \cite*{Dottori03} estimated the corotation radius position using Fourier analysis.
\item

\item \noindent {\it NGC2903}: There are two evaluations of the corotation radius for this galaxy, coincident with each other and with the metallicity break found at about 5 kpc using the polynomials of order 3, 4 and 5, in addition to the two connected lines. The metallicity distribution is well sampled by 36 \hii regions \citep{Pilyugin02} and the results seem to be compatible with the spiral structure extension. The same result is obtained using the \cite{Zaritsky94}, which make this galaxy a strong point in the correlation presented in Figure \ref{fig4}. We cannot exclude the possibility of the corotation being at 9 kpc, where our polynomial fits presents an inflexion, but there are no references in the literature supporting that. This would put the corotation radius at the extremity of two well behaved symmetrical spiral arms, like it happens in other galaxies, and would bring the pattern speed to 31 km/s/kpc, a more usual one than 57 km/s/kpc.
\item

\item \noindent {\it NGC3031 (M81)}: The estimates of the corotation radius for this galaxy  is sometimes mentioned to be around 8 kpc and other times around 14 kpc, as can be seen in the work by \cite{Kendall+08}, recalibrated according the projection parameters assumed in this work. In spite of this, based on its frequency diagram, the extension of its spiral structure favors a spiral pattern speed of 20 km/s/kpc. This speed results in a corotation radius near 8 kpc, compatible with the breaks found using the polynomial fits of order 3 to 5 and with the two straight line fits of the radial abundances determined by \cite{Stanghellini+10} using planetary nebulae. These results are also consistent with the breaks found using the radial abundance distribution by \cite{Patterson+12} for \hii regions with {\bf the calibrator KK04, given in their Figure 9, which also includes data from other authors}. For this reason this galaxy is assumed as a strong point in our correlation.
\item

\item \noindent {\it NGC3319}: This is another example of a highly inclined galaxy. In contrast with NGC0224 it has well defined break identified by all models that we fitted to its metallicity distribution. Unfortunately it is sampled only by 13 \hii regions for which only one calibration method was applied \citep{Zaritsky94}, so it is represented by a weak point in the correlation in Figure \ref{fig4}. Its rotation curve is almost a rigid body rotation curve (according the results in optic by \citealt{Garrido02} and in radio by \citealt{Moore98}). Since its spiral pattern speed and the angular velocity associated to the rotation curve are similar along a significant part of the disk it may explain the "ring" with less \hii regions in this galaxy. Note that the spiral pattern speed mentioned in the literature is compatible with the spiral structure extension.
\item

\item \noindent {\it NGC3359}: Galaxy with 94 \hii regions observed by \cite{Martin95} and \cite{Zahid2011}. In both references the authors found breaks in the radial metallicity distribution using different calibration methods and associated it to a possible effect of the corotation radius. \citeauthor{Zahid2011} confirmed the statistical significance of the flattening using the an F-test. Similarly, according our procedure, polynomials of order 4 and 5,  two connected lines and two straight lines with a step fit the data better than the linear model. For all this cases a break in the metallicity distribution is found around 7.9 kpc, near the corotation radius determined by \cite{Rozas08}. For these reasons this galaxy is represented as a strong point in the correlation shown in Figure \ref{fig4}.
\item

\item \noindent {\it NGC3621}: The radial oxygen abundance of this galaxy was extensively studied by \cite*{Bresolin12} using different calibrators for its 73 \hii regions. These authors found a significant flattening in the metallicity gradient of this galaxy after R25 using the same procedure of \cite{Zahid2011}. Our procedure agrees with their results and two connected lines and two straight lines with a step fit the data and exhibit an evident break near 8.7 kpc. These results from different analyses turn this galaxy a strong point in the correlation shown in Figure 4. Unfortunately there are not many studies on the corotation radius of this galaxy. \cite{Tamburro08} determined its spiral pattern speed by deriving geometrically the timescale of the star formation driven by its spiral arms. For this purpose they used SINGS and THINGS data sets to verify the angular offsets between the 24 $\mu$m and the \hi emission tracers. The original result of this work points to a corotation radius of 2.2 kpc. However these authors tag this value as a result of a bad quality fit. We derive a better value (6.2 kpc) by discarding the most discrepant data from the analysis by \cite{Tamburro08} and comparing the results with the observed rotation curve by \cite{Blok08}.
\item

\item \noindent {\it NGC4254 (M99)}: Galaxy sampled by 18 \hii regions with oxygen abundance determination listed by \cite{Moustakas+10}. The objects are relatively well distributed along the disk but the observations do not go too far from the median corotation radius mentioned in the literature. None of these references on the corotation fully agree with each other, but the metallicity break found using a 3rd order polynomial and two lines with a step on the data calibrated with the PT05 method are compatible with the median results by \cite{Elmegreen92}, \cite{Kranz01} and \cite{Gonzalez96}, which we adopted in our correlation. On the other hand, the break radius found using the KK04 calibration and two connected lines fit is consistent with the corotation radius determined by \cite{Gonzalez96} and we cannot discard their value as the real position of the corotation radius. Since the results from different calibrations do agree with the presence of a break, but they do not agree with a specific position in the radial oxygen distribution, this galaxy is represented as a weak point in our correlation.
\item

\item \noindent {\it NGC4321 (M100)}: This is the most explored galaxy regarding the evaluation of the corotation radius and the spiral pattern speed. A histogram of the corotation radii taken from the literature reveals a double peaked distribution, indicating that two spiral pattern modes could be superimposed. The extension of the spiral structure is compatible with the corotation radius of the external peak, which is also consistent with the metallicity breaks found with 4th and 5th order polynomials. Unfortunately the abundance distribution of this galaxy is sampled by only 10 \hii regions, so we had to discard it from our analysis.
\item

\item \noindent {\it NGC4736 (M94)}: Galaxy with peculiar arms very extended in relation the brighter part of the disk where a ring can be observed. The sample of \hii regions consist in only 8 objects, all of them very concentrated in the internal regions of the galaxy, and for this reason this galaxy was not included in our analysis.
\item

\item \noindent {\it NGC4826}: Another galaxy with few \hii regions (8) studied along the galactic disk. The objects are very concentrated at galactic radii internal to the expected position for the corotation radius. Therefore this object was also excluded of our analysis.
\item

\item \noindent {\it NGC5033}: This galaxy presents an asymmetric spiral structure with distinct pitch angles. Considering that there is only 8 \hii regions observed along this galaxy we did not take it into account in our correlation analysis.
\item

\item \noindent {\it NGC5055 (M63)}: Flocculent spiral galaxy with only five \hii regions studied in terms of metallicity. Hence, it was not possible to perform a reliable study of its metallicity distribution, and consequentely we discarded it from our correlation study.
\item

\item \noindent {\it NGC5194 (M51)}: Grand-design spiral galaxy with 16 studies relative to its corotation radius. The values are bimodally distributed around 5.5 and 11 kpc (near the outer limit of the spiral structure). There is a very precise study on the metallicity distribution by \cite*{Bresolin04}, using temperature-sensitive line methods in 10 \hii regions and also an extensive study by \cite{Moustakas+10}, using the KK04 and PT05 methods in 32 \hii regions compiled in the literature. In spite of the small amount of \hii regions in the first case, breaks in the metallicity distribution were found using polynomials of order 3 and 4, with values that are coincident with the median corotation radii from the literature. Breaks at the same position were found in the second case using fits of 3rd order polynomial, two connected lines and two straight lines with a step. Using the same method of \cite{Mishurov02}, \cite{Acharova08} determined the spiral pattern speed, of which the corresponding corotation radius is coincident with the radius of the break. For these reasons this galaxy is considered a strong point in the correlation shown in Figure \ref{fig4}.
\item

\item \noindent {\it NGC5236 (M83)}: This galaxy is possibly one of the best cases of the present study. \cite{Bresolin09} obtained O abundances for 49 \hii regions, with low scattering, which combined with data from 23 \hii regions previously observed shows a clear change of slope associated to a step at 5.5 arcmin (8 kpc).  There are very discrepant determinations of the corotation radius for this object. To make a choice, we must first examine some aspects of the morphology of this galaxy. 1-) The rotation curve is flat to large distances, which indicates that the galaxy is not disturbed by any strong external influence; 2-) A seminal paper of \citeauthor{Elmegreen95} \citeyearpar[hereafter  E\&E]{Elmegreen95} shows that almost all spiral galaxies have a well defined 2-arms structure up to a distance of 0.5 R25. We see the end of this structure at about 3 arcmin from the centre (also determined by E\&E for this galaxy), which places R25 and the corotation radius at about 6 arcmin from the center. 3-) A map of the gas distribution \citep{Crosthwaite+02} shows a ring void of gas with radius 5.5 arcmin, which resembles the one which exists at corotation, in our Galaxy. As discussed by \cite{Amores2009}, such a gap is not only observed in our Galaxy but also predicted by theory as an effect of corotation. In general the corotation radius mentioned in the literature is relatively smaller than the indicators mentioned here. A same group of researchers (\citealt{Talbot79} and \citealt{Jensen81}) obtained two discrepant values of $\Omega_p$ (22 and 53 km/s/kpc, recalibrated accoriding our projection parameters) using different methods. In their 1981 paper, these authors analyse the reasons for this discrepancy and state their preference for the largest value (which implies a smaller corotation radius). They comment that the first method relies on the fact that the star formation rate is proportional to ($\Omega-\Omega_p$) and reject this idea, although it gives a convincing fit of the data. Note that this proportionality is precisely the one we verify in our Galaxy. On the other hand, the argument that they present in favour of the largest $\Omega_p$ is the absence of a significant drift of the young stars with respect to the arms, near the inner end of the arms. They consider that the drift should be proportional to ($\Omega-\Omega_p$); if the drift is very small, the observed region must be close to corotation. However they do not take into account the mechanism described by \cite{Lord91} for M83 in which the gas gets trapped in the bottom of the channel-shaped potential traced by the arms, and flows along them. Consequently, the newborn stars have initial velocities along the arms, so no drift would be observed. Therefore we adopt 22.5 km/s/kpc as the correct solution. This places the corotation at 8.1 kpc, in agreement with our previous arguments.
\item

\item \noindent {\it NGC5457 (M101)}: This galaxy has a well sampled metallicity distribution \cite{Pilyugin03} with 76 \hii regions, which presents a clear flattening in the external regions. All the models that we use (polynomials, 2 lines, or two lines with a step) give a better fit than a single straight line, and point to a break at 19 kpc. However, in his 2003 paper, Pilyugin questions the existence of the bends in the slopes of the oxygen abundance in spiral galaxies. The best data for NGC5457, analysed with Pilyugin's calibration method, is shown in his figure 1b, where we can see that all the points have abundances larger than 12 + log(O/H) = 8.2.  In the same paper it is said that Pilyugin's formula is valid above this limit (8.2). If we take into account the uncertainties and the data dispersion (0.2 dex) we see that in spite of many data points being close to the confidence limit they cannot be discarded. \cite{Kennicutt03} presents a set of measurement of 20 points, determined using the "direct" calibration, for which the radial oxygen distribution is significantly less flat than in the previous case.  For these data, fitting with 2 lines, or with 2 lines with a step, the results are better than with a single line, and give breaks at 17.7 and 17.9 kpc respectively. The median value of these breaks are in relative agreement with the corotation radius found using the results by \cite{Roberts75}, which is also compatible with the extension of the spiral structure. This last argument forces us discard the values of the spiral pattern speed by \cite{Meidt05} and \cite{Sheth02}. Since we found two sets of data in relative agreement and based on different calibration methods, this galaxy is considered as a strong point.
\item

\item \noindent {\it NGC6946}: Galaxy sampled by only 7 \hii regions compiled by \cite{Moustakas+10} and for this reason it will not included in our correlation analysis.
\item

\item \noindent {\it NGC7479}:  This object is also one of the most extensively studied in term of its corotation radius, with almost 20 references on it. Most of the corotation radii are lower than 11 kpc, all of them incompatible with the spiral structure extension, except for the results by \cite{Courtes71} and \cite{Fathi09}. In terms of radial oxygen abundance the data is almost exclusively from the observations by \cite*{Martin00} of 69 \hii regions candidates, from which at least 63 were suitable for abundances studies with four different calibrations methods. They all look similar, which suggests that the step seen in the radial distribution is not an effect of a calibration error. Analyzing the distribution with the smallest scattering in the data, we find that two lines with a step at 17 kpc and two connected lines with break at 16.2 kpc give better fits than a single line. Considering that the mean metallicity break radii obtained from different calibrations methods are similar, this galaxy is considered as a strong point in the correlation of Figure \ref{fig4}.

\end{description}

\section{Discussion}

\par The existence of a minimum of metallicity at corotation, and the corresponding rising metallicity in the external parts of a number of spiral galaxies, can be understood under one of the simplest assumptions that can be made, namely that the star formation rate (SFR) is proportional to $\mu R \mid \Omega-\Omega_p \mid $, or similar expressions, eg. \citet{Mishurov02}. This is the surface density of gas $\mu$ times the velocity at which the gas penetrates the spiral arms, or in other words, the rate at which the gas feeds the star-formation machine. A similar recipe for star formation was proposed by \citet{WyseSilk89}, but they did not give importance to the fact that one must use the modulus of the frequency difference. The relative velocity of the gas of the disk with respect to the spiral pattern is zero at corotation, and reverses its direction at larger radii. The oxygen abundance at a given radius is proportional to the integral of star-formation rate with time, since O is mostly contributed by type II supernovae. These stars are short lived and do not have time to migrate; almost as soon as they are born they deliver new oxygen into the interstellar medium. Beyond the corotation the side of the arms on which the gas penetrates is reversed, but the modulus $\mid \Omega-\Omega_p \mid $ increases again, and consequently also the rate of enrichment of the medium in oxygen.

This idea of proportionality between the amount of gas which enters the arms and the number of stars formed is specially easy to accept if we adopt the classical description of spiral arms of \cite{Kalnajs1973}, in which the arms are due to crowding of stellar orbits in some regions of the galactic disk. Therefore, the arms are "massive" entities, and behave like elongated potential wells or grooves in the axisymmetric potential. The interstellar gas gets trapped and accumulates into the elongated wells until the Jeans mass is reached and star formation starts. In this process the flux of gas reaching an arm is in some way "integrated" over the time. The above explanation gives us a rough picture of processes which might be going on. The description is a little oversimplified, however, since it is not only the velocity of galactic rotation relative to the spiral pattern that forces the gas to penetrate into the arms. The gravity of the arms, by itself, is able to attract the gas.

The presence of a step in metallicity at corotation, in addition to the change of slope, as we observed in several galaxies, requires another explanation. In our view, this is a consequence of the fact that the two sides of corotation are isolated one from the other, with the radial gas flow induced by the spiral pattern going inwards inside corotation and outwards on the outer side. As a consequence, the two sides have no exchange of gas and evolve independently. This interpretation of the step was discussed by L+8 in the case our Galaxy. Such opposite flows are not only predicted by the theory of spiral arms \cite*{Mishurov2009}, but confirmed by hydrodynamic simulations (eg \cite{L'epine2001}, Figure \ref{fig4}) and directly observed, like in the case of NGC1365 \cite*{Elmegreen2009}. In addition to be isolated one from the other, the explanation requires that the two regions have different SFR. This is easy to accept since we always see the inner regions of galaxies having higher stellar and gas densities.

Both the positive gradients in external regions of spiral galaxies, and the metallicity steps at corotation radius, of which we confirm the existence in this work, have an important consequence for the understanding of the evolution of spiral structures. A relatively long time is required to build up these features. This is a strong argument against the concept of "transient" spiral arms. If the corotation radius did not remain for a long time at a same radius, the discontinuities in chemical abundance would be smoothed out. L+8 estimated the minimum lifetime of the present spiral structure of our Galaxy to be about 3 billion years. This is also a typical number for the galaxies investigated here, based on the same arguments.

The correlations presented in Figures \ref{fig4} in some way validate the methods used to find the corotation radius in the literature, and the method that we used to locate breaks in the metallicity gradients. If any one of these methods failed frequently, the correlations would be much worse. This is also an argument against the idea that galaxies often have multiple pattern speeds. If this were so, we could have the corotation radius of one of the pattern speeds confronted with the metallicity features of another one, and this would destroy the correlation. It seems that at least a long lived dominant pattern speed exists for each galaxy. On the other hand we cannot discard the possibility of some mis-association between breaks in the radial metallicity distribution and the other resonances, specially the Lindblad resonances and the 4:1 resonance.

\begin{figure}
\includegraphics[width=84mm]{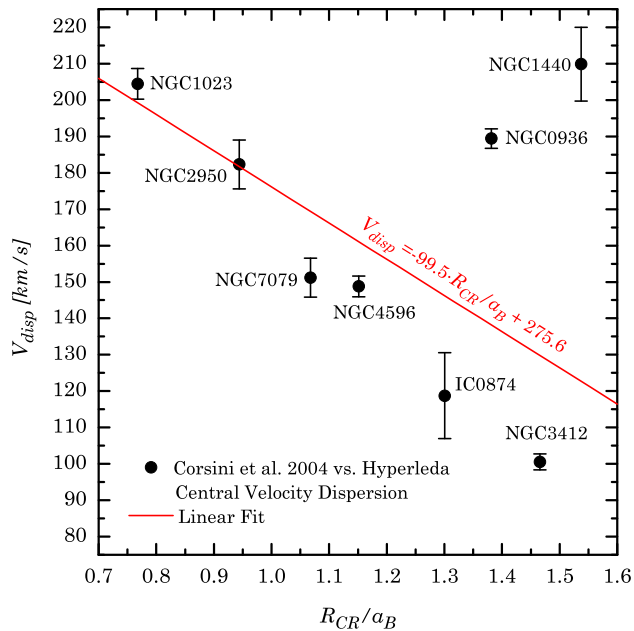} \caption{Correlation between the corotation radius($R_{CR}$) scaled by semi-major axis length ($a_{B}$) and the central velocity dispersion ($V_{disp}$) of the galaxies studied by Corsini et al (2004) with the TW-method.}
 \label{fig6}
\end{figure}

It is worthwhile, at this point, to pay some attention to minor deviations with respect to the one-to one relationship. One may note a tendency of the metallicity break being at a slightly (10 \%) larger radius than corotation, in Figure \ref{fig4}a. In agreement with this tendency, the fitted straight lines to the angular speeds (Figure \ref{fig4}b) have a slight bias in the sense of spiral pattern speeds being faster than the one-to-one match line. The minor deviations of the two correlations could be attributed, for instance, to a systematic offset of the metallicity break towards a radius larger than corotation. The position of the minimum of the star formation rate (which originates the minimum of metallicity that we detect) can be influenced by the gradient of gas density. One can note for instance in the result of a hydrodynamic simulation of the effect of the spiral potential perturbation on the gas density, in Figure 4 of \citet{L'epine2001} paper, that the resulting minimum of gas density is at a slightly larger radius than corotation. On the other hand, with respect the spiral pattern speed,  the deviations with respect to perfect match could be attributed to systematic errors in the determination of the corotation radii, specially when the Tremaine-Weinberg method \citep[TW-method hereafter]{TW84} is applied. Apparently the TW-method provides pattern speeds systematically higher than the other methods, especially when it is applied to bars. If the spiral arms and the bar have the same pattern speed, as suggested by the clear bar-arm connection in most galaxies (see discussions in \citealt{Kormendy79} and \citealt{Block04}), there could be a source of errors associated to the TW-method which results in higher velocities as seem for the objects with higher spiral pattern speed in \ref{fig4}. Possibly the TW-method is affected by the dispersion of the stellar velocities. This suspicion is supported by the results of an analysis of the corotation radius scaled by the semi-major axis obtained by \cite*{Corsini04} using the TW-method, compared to the central velocity dispersion compiled from the HYPERLEDA\footnote{http://leda.univ-lyon1.fr} database. Figure \ref{fig6} shows a clear tendency of larger velocity dispersions being associated with faster pattern speeds (or equivalently, with smaller corotation radii). Large velocity dispersions introduce extra velocity components in the velocity measurements integrated along the line-of-sight. Since in the TW-method the line of sight velocity is supposed to belong to the galactic plane, the extra component introduces a systematic error source. The systematic errors associated with the TW method are discussed in detail by \cite{Meidt2008}.

\section{Conclusions}

In this work we showed the existence of a clear correlation between breaks in the metallicity distribution and corotation radii of spiral galaxies. To achieve this task we compiled all the references that we could find on rotation curves, metallicity distributions, and spiral pattern speeds or corotation radii of galaxies. The aim was to construct a sample of galaxies for which there is at least one measurement in the literature for each one of these quantities. The references are presented in tables in the Appendix. Since the data from the literature spans many years, different authors, and different methods, we established criteria to standardize the data, in order to avoid combining data that were obtained based on discrepant rotation curves and/or galactic distances.

To locate minima, inflexions or breaks in the metallicity distribution, we developed an automatic procedure which consists in fitting, in sequence, several low order polynomials, a pair of straight connected lines forming an angle, and a pair of straight lines with a step at the connection between them. The significance of each of these fits was evaluated using the chi-squared by degree of freedom. Coincident values for minima, inflexions and breaks were registered. Almost all galaxies that were better fitted by other models than the linear fit presented coincident minima for different curves (different polynomial order or the set of two lines), which means that the minimum is not an artefact of a given fit.

The distribution of spiral pattern speeds presents a small concentration around 20 km/s/kpc, so that the spiral pattern speed of our Galaxy can be considered typical in the context of our sample. The distribution of corotation radii, on its turn, presents a cutoff around 15 kpc, which is consistent with the spiral density wave theory, since typical large spiral galaxies have a radius of about 30 kpc. The outer Lindblad resonance is supposed to be about the boundary of the visible disks and corotation is expected to be about midway between the Lindblad resonances. We noted a minor systematic deviation (about 10\%) of the correlation between metallicity breaks and corotation from the one-to-one relation, in the sense that the metallicity breaks are at a larger radius. This could be due to a real shift of the break with respect to corotation, or to a slight error in the determination of the corotation radius, possibly associated with the TW-method.

The existence of a minimum of metallicity at corotation, or  equivalently, of a raising gradient in external regions of a galaxy, as we found in many spiral galaxies presented here, can be explained by a star-formation recipe like SFR $\propto ~\mu R \mid \Omega-\Omega_p \mid $ or similar ones. We argued in the previous section that this is a quite reasonable hypothesis. The presence of steps in the metallicity distribution is a little more difficult to elucidate. Our hypothesis is that it is a consequence of the radial flows of gas going in opposite directions on each side of corotation, which tend to produce a depletion of gas at this resonance. This creates like a barrier, with no transfer of gas from one side to the other, so that the two sides evolve independently. We do not discard the effect of other resonances.

An important consequence of the existence of both the breaks in metallicity gradients and of metallicity steps is that they points towards a long term stability of the corotation radius of the host galaxies. A varying corotation radius would smooth out such features. In other words the spiral structures of many galaxies are long-lived, as discussed by L+8 in the case of our Galaxy, for which a minimum age of the spiral structure was estimated to be 3 billion years. Our result also favours the idea that each galaxy has a single, or at least a quite dominant pattern speed, otherwise the correlation that we found would be destroyed.

\section{Acknowledgements}
This work was supported by the Sao Paulo State Agency FAPESP through the grant 09/05181-8. Besides, it has been benefited by the NASA's Astrophysics Data System (ADS) and the Extra-galactic Database (NED) services. Their open software used in this research is greatly acknowledged. This work made use of THINGS, 'The \hi Nearby Galaxy Survey' \citep{Walter2008}. We acknowledge the usage of the HyperLeda database (http://leda.univ-lyon1.fr). Finally, we would like to thank the anonymous referee for his numerous and valuable suggestions for improvements of this paper.

\bibliographystyle{mn2e}
\bibliography{metalgradcorot5}

\appendix
\section{References in the Literature}

In Table \ref{tbl-a1} we compile the references on corotation radii, spiral pattern speeds, metallicity distributions and rotation curves used to obtain the results in Tables \ref{tbl-1}, \ref{tbl-2} and \ref{tbl-a2}. The details on the references consulted about the corotation radii, with standardized values for each reference can be found in Table \ref{tbl-a2}.

\begin{table*}
 \caption{Numerical index for each reference mentioned in Tables \ref{tbl-1}, \ref{tbl-2} and \ref{tbl-a2}.}
 \label{tbl-a1}
 \centering
\begin{tabular}{rlrlrl}
 \hline
  \hline
{\it {\bf ID}} &      {\it {\bf References}} &  {\it {\bf ID}} &  {\it {\bf References}} &  {\it {\bf ID}} &  {\it {\bf References}} \\
 \hline
  \hline

 {\bf [1]} &           \cite{Aguerri96}  &   {\bf [47]} &    \cite{Garnett97}  &   {\bf [93]} &       \cite{Quillen95}  \\
 {\bf [2]} &            \cite{Aguerri00} &   {\bf [48]} &   \cite{Garrido02}   &   {\bf [94]} &          \cite{Rand04}  \\
 {\bf [3]} &               \cite{Ball92} &   {\bf [49]} &    \cite{Garrido05}  &   {\bf [95]} &          \cite{Rand95}  \\
 {\bf [4]} &            \cite{Beckman90} &   {\bf [50]} &       \cite{Gerin91} &   {\bf [96]} &        \cite{Roberts75} \\
 {\bf [5]} &         \cite{BelleyRoy92}  &   {\bf [51]} &    \cite{Gonzalez96} &   {\bf [97]} &           \cite{Rots75} \\
 {\bf [6]} &              \cite{Bland97} &   {\bf [52]} &   \cite{Gottesman75} &   {\bf [98]} &      \cite{RoyWalsh97}  \\
 {\bf [7]} &              \cite{Blok08}  &   {\bf [53]} &      \cite{Helfer95} &   {\bf [99]} &          \cite{Rozas08} \\
 {\bf [8]} &         \cite{Bresolin04}   &   {\bf [54]} &       \cite{Henry95} &  {\bf [100]} &          \cite{Saha02}  \\
 {\bf [9]} &          \cite{Bresolin+05} &   {\bf [55]} &   \cite{Hernandez05} &  {\bf [101]} &         \cite{Sakai99}  \\
{\bf [10]} &           \cite{Bresolin12} &   {\bf [56]} &      \cite{Jensen81} &  {\bf [102]} &       \cite{Sakhibov87} \\
{\bf [11]} &           \cite{Bresolin09} &   {\bf [57]} &   \cite{Jorsater95}  &  {\bf [103]} &          \cite{Salo00}  \\
{\bf [12]} &           \cite{Bresolin12} &   {\bf [58]} &    \cite{Kamphuis92} &  {\bf [104]} &    \cite{Schinnerer00}  \\
{\bf [13]} &          \cite{ButaZhang09} &   {\bf [59]} &     \cite{Kelson96}  &  {\bf [105]} &      \cite{Sempere95a}  \\
{\bf [14]} &            \cite{Canzian97} &   {\bf [60]} &     \cite{Kendall98} &  {\bf [106]} &     \citet{Sempere95b}  \\
{\bf [15]} &            \cite{Canzian93} &   {\bf [61]} &   \cite{Kennicutt96} &  {\bf [107]} &        \cite{Shanks97}  \\
{\bf [16]} &           \cite{Canzian+93} &   {\bf [62]} &   \cite{Kennicutt93} &  {\bf [108]} &         \cite{Sheth02}  \\
{\bf [17]} &        \cite{Castellanos02} &   {\bf [63]} &     \cite{Knapen00}  &  {\bf [109]} &       \cite{Shields91}  \\
{\bf [18]} &             \cite{Cedres02} &   {\bf [64]} &     \cite{Kranz01}   &  {\bf [110]} &            \cite{Shu71} \\
{\bf [19]} &               \cite{Cepa90} &   {\bf [65]} &       \cite{Laine96} &  {\bf [111]} &    \cite{Silbermann99}  \\
{\bf [20]} &            \cite{Corsini04} &   {\bf [66]} &     \cite{Laine+98}  &  {\bf [112]} &       \cite{Skillman96} \\
{\bf [21]} &            \cite{Courtes71} &   {\bf [67]} &        \cite{Lee02}  &  {\bf [113]} &        \cite{Sofue99}   \\
{\bf [22]} &           \cite{Crockett06} &   {\bf [68]} &   \cite{Lindblad96}  &  {\bf [114]} &  \cite{Stanghellini+10} \\
{\bf [23]} &            \cite{Daigle06}  &   {\bf [69]} &       \cite{Lord91}  &  {\bf [115]} &        \cite{Talbot79}  \\
{\bf [24]} &                 \cite{RC3}  &   {\bf [70]} &        \cite{Lowe94} &  {\bf [116]} &       \cite{Tamburro08} \\
{\bf [25]} &             \cite{delRio98} &   {\bf [71]} &   \cite{Magrini2007} &  {\bf [117]} &          \cite{Thim03}  \\
{\bf [26]} &           \cite{Dessart08}  &   {\bf [72]} &    \cite{Marquez02}  &  {\bf [118]} &        \cite{Toonen08}  \\
{\bf [27]} &               \cite{Diaz07} &   {\bf [73]} &      \cite{Martin95} &  {\bf [119]} &         \cite{Tully74}  \\
{\bf [28]} &       \cite{DorsCoppetti05} &   {\bf [74]} &     \cite{Martin00}  &  {\bf [120]} &        \cite{vanZee99}  \\
{\bf [29]} &              \cite{Duval85} &   {\bf [75]} &     \cite{Martin97}  &  {\bf [121]} &        \cite{vanZee98}  \\
{\bf [30]} &          \cite{Elmegreen85} &   {\bf [76]} &      \cite{McCall85} &  {\bf [122]} &       \cite{vanZee+98}  \\
{\bf [31]} &          \cite{Elmegreen90} &   {\bf [77]} &      \cite{Meidt05}  &  {\bf [123]} &      \cite{Dottori03}   \\
{\bf [32]} &         \cite{Elmegreen95}  &   {\bf [78]} & \cite{Moellenhoff95} &  {\bf [124]} &         \cite{Vera01}   \\
{\bf [33]} &          \cite{Elmegreen89} &   {\bf [79]} &      \cite{Moore98}  &  {\bf [125]} &     \cite{VilaCosta92}  \\
{\bf [34]} &          \cite{Elmegreen92} &   {\bf [80]} &  \cite{Moustakas+10} &  {\bf [126]} &     \cite{Vilardell07}  \\
{\bf [35]} &        \cite{Elmegreen2009} &   {\bf [81]} &      \cite{Munoz04}  &  {\bf [127]} &         \cite{Vinko04}  \\
{\bf [36]} &              \cite{Evans87} &   {\bf [82]} &      \cite{Newton80} &  {\bf [128]} &         \cite{Vinko06}  \\
{\bf [37]} &              \cite{Fathi09} &   {\bf [83]} &         \cite{Oey93} &  {\bf [129]} &     \cite{Visser1980b}  \\
{\bf [38]} &          \cite{Ferguson98}  &   {\bf [84]} &         \cite{Oey03} &  {\bf [130]} &         \cite{Warner73} \\
{\bf [39]} &         \cite{Ferrarese96}  &   {\bf [85]} &  \cite{Patterson+12} &  {\bf [131]} &      \cite{Westpfahl98} \\
{\bf [40]} &            \cite{Fierro86}  &   {\bf [86]} &      \cite{Pence90}  &  {\bf [132]} &          \cite{Wong00}  \\
{\bf [41]} &          \cite{Freedman94}  &   {\bf [87]} &   \cite{Pilyugin03}  &  {\bf [133]} &       \cite{Worthey+05} \\
{\bf [42]} &          \cite{Freedman01}  &   {\bf [88]} &   \cite{Pilyugin02}  &  {\bf [134]} &        \cite{Zahid2011} \\
{\bf [43]} &     \cite{Garcia-Burillo94} &   {\bf [89]} &   \cite{Pilyugin04}  &  {\bf [135]} &       \cite{Zaritsky94} \\
{\bf [44]} &     \cite{Garcia-Burillo98} &   {\bf [90]} &     \cite{Puerari92} &  {\bf [136]} &        \cite{Zimmer04}  \\
{\bf [45]} & \citet{Garcia-Burillo1993a} &   {\bf [91]} &    \cite{Puerari97}  &  {\bf [137]} &      \cite{Rosales+11}  \\
{\bf [46]} & \citet{Garcia-Burillo1993b} &   {\bf [92]} &    \cite{Puerari93}  &              &                         \\
 \hline
  \hline
\end{tabular}

\end{table*}

\begin{table*}
 \caption{Reference indexes for the corotation radii (R$_{CR}$) of each galaxy in our sample. Each index is preceded by the values found in the papers and re-scaled according the procedures in Section 3 and the Table \ref{tbl-1}. The symbol "$\vee$" indicates two possible values and the symbol "-" separate a range of values.}
 \label{tbl-a2}
 \centering
\begin{tabular}{ll}
\hline \hline
                    {\it {\bf GALAXY}} &   {\it {\bf R$_{CR}$ [kpc ]}} \\
\hline \hline
                    {\bf IC0342}       &   5.8 {\bf [34]}; 14.6 {\bf [125]}                                                                                                                                                     \\ \hline
                    {\bf NGC0224}      &   14.8 {\bf [125]}; 16.4 {\bf [96]}                                                                                                                                                    \\ \hline
                    {\bf NGC0598}      &   3.4 {\bf [96]}; 1.2 $\vee$ 7.0 {\bf [82]};  5.9 {\bf [125]}; 2.9 {\bf [34]}; 4.3 {\bf [92]};  4.0 {\bf [21]}; 4.7 {\bf [130]}                                                        \\ \hline
                    {\bf NGC0628}      &   4.8 {\bf [19]}; 4.6 {\bf [34]}; 4.6 {\bf [125]}; 2.2 $\vee$ 2.9 {\bf [90]}                                                                                                           \\ \hline
                    {\bf NGC1068}      &   3.1 $\vee$ 3.3 {\bf [94]}; 1.4 {\bf [53]}                                                                                                                                            \\ \hline
                    {\bf NGC1232}      &   12.6 {\bf [34]}                                                                                                                                                                      \\ \hline
                    {\bf NGC1365}      &   13.9 {\bf [68]}; 12.3 {\bf [57]}; 14.1 {\bf [35]}; 10 $\vee$ 20.7 {\bf [124]}                                                                                                        \\ \hline
                    {\bf NGC1530}      &   13.4 {\bf [124]}                                                                                                                                                                     \\ \hline
                    {\bf NGC1566}      &   8.8 {\bf [23]}                                                                                                                                                                       \\ \hline
                    {\bf NGC2403}      &   5.1 {\bf [125]}                                                                                                                                                                      \\ \hline
                    {\bf NGC2543}      &   4.4 $\vee$ 7.5 {\bf [124]}                                                                                                                                                           \\ \hline
                    {\bf NGC2903}      &   4.2 {\bf [108]}; 4.0 {\bf [125]}                                                                                                                                                     \\ \hline
\multicolumn{ 1}{r}{{\bf NGC3031}}     &   13.1 {\bf [129]}; 9.4 $\vee$ 8.1 {\bf [34]}; 11.9 $\vee$ 12.9 {\bf [52]};  9.1 $\vee$ 9.8 {\bf [70]}; 11 {\bf [96]}; 11.6 {\bf [97]}; 12.4 {\bf [102]}; 9.9 $\vee$ 15.5 {\bf [110]}; \\
\multicolumn{ 1}{r}{{\bf        }}     &   8.5 {\bf [116]}; 9.5 {\bf
[131]}; 12.3 \bf{\bf [60]}
\\ \hline
                    {\bf NGC3319}      &   4.9 - 7.4 {\bf [32]}; 8.7 {\bf [13]}                                                                                                                                                 \\ \hline
                    {\bf NGC3359}      &   5.1 {\bf [3]}; 6.6 {\bf [99]}; 6.1 {\bf [135]}; 6.2 {\bf [73]}                                                                                                                       \\ \hline
                    {\bf NGC3621}      &   6.2 {\bf [116]}                                                                                                                                                                      \\ \hline
                    {\bf NGC4254}      &   13.6 {\bf [34]}; 12.1 {\bf [64]}; 16.5 {\bf [51]}                                                                                                                                    \\ \hline
\multicolumn{ 1}{l}{{\bf NGC4321}}     &   8.9 - 7.6 $\vee$ 6.4-8.9 {\bf [106]}; 7.3 {\bf [20]}; 8.8 {\bf [33]}; 6.3 - 7.2 {\bf [95]}; 7.6 $\vee$ 6.8 $\vee$ 8.4 {\bf [14]}; 8.6 $\vee$ 14 {\bf [43]};                          \\
\multicolumn{ 1}{l}{{\bf        }}     &   7.9 - 10.0 {\bf [15]}; 9.2 {\bf [34]}; 12 {\bf [84]}; 6.2 {\bf [18]}; 3.1 - 4.5 {\bf [106]}; 7 {\bf [108]}; 9.6 $\vee$ 10.9 {\bf [94]}; 1.3 $\vee$ 11.3 {\bf [44]};                  \\
\multicolumn{ 1}{l}{{\bf        }}     &   3.3 {\bf [63]}; 4.2 $\vee$ 8.6 $\vee$ 13.1 {\bf [55]}                                                                                                                                \\ \hline
                    {\bf NGC4736}      &   3.1 {\bf [50]}; 0.6 {\bf [132]}; 0.7 $\vee$ 0.8 {\bf [78]}                                                                                                                           \\ \hline
                    {\bf NGC4826}      &   1.1 $\vee$ 1.4 {\bf [94]}                                                                                                                                                            \\ \hline
                    {\bf NGC5033}      &   3.2 {\bf [77]}                                                                                                                                                                       \\ \hline
                    {\bf NGC5055}      &   12.8 {\bf [34]}; 14 {\bf [78]}; 6.7 {\bf [77]}; 15.4 {\bf [96]}                                                                                                                      \\ \hline
\multicolumn{ 1}{l}{{\bf NGC5194}}     &   5.7$\vee$ 10.1 {\bf [33]}; 5.4 $\vee$ 10.2 {\bf [96]}; 2.8 {\bf [136]}; 5.34 $\vee$ 6.7 {\bf [119]}; 6.9 {\bf [46]}; 2.8 - 5.6 {\bf [103]}; 5.8 {\bf [34]};                          \\
\multicolumn{ 1}{l}{{\bf        }}     &   10.1 {\bf [125]}; 13.2 {\bf [84]}; 10.2 {\bf [110]}; 5.4 {\bf [116]}                                                                                                                 \\ \hline
                    {\bf NGC5236}      &   3.8 {\bf [136]}; 3.4 {\bf [69]}; 3.4 {\bf [56]}; 8.1 {\bf [115]}; 4.1 {\bf [125]}                                                                                                    \\ \hline
                    {\bf NGC5457}      &   1.2 {\bf [108]}; 1.7 {\bf [77]}; 16.7 {\bf [96]}                                                                                                                                     \\ \hline
                    {\bf NGC6946}      &   2.8 {\bf [34]}; 4.6 {\bf [125]}; 1.9 {\bf [108]}; 3.7 {\bf [136]}; 4.9 $\vee$ 9.7 {\bf [118]};                                                                                       \\ \hline
\multicolumn{ 1}{l}{{\bf NGC7479}}     &   3.0 $\vee$ 4.7 {\bf [123]}; 10.7 {\bf [2]}; 9.5 {\bf [4]}; 9.5 {\bf [29]}; 12.5 $\vee$ 17.8 {\bf [21]};                                                                              \\
\multicolumn{ 1}{l}{{\bf        }}     &   7.6 {\bf [93]}; 14.4 {\bf [25]}; 9.6 {\bf [105]}; 9.3 {\bf [65]}; 9.3 {\bf [91]}; 0.5 {\bf [40]}; 14.6 {\bf [37]}; 10.0 {\bf [13]}                                                   \\
\hline \hline
\end{tabular}
\end{table*}

\end{document}